\documentclass[aps,nopacs,twocolumn,preprintnumbers,amsmath,amssymb,longbibliography]{revtex4}
\usepackage{epsfig}
\usepackage{graphicx}% Include figure files
\usepackage{dcolumn}% Align table columns on decimal point
\usepackage{color}
\usepackage{bm}% bold math

\begin{document}
\title{Graphene Plasmonics for Terahertz to Mid-Infrared Applications}
\author{Tony Low}
\author{Phaedon Avouris}
\affiliation{IBM T.J. Watson Research Center, 1101 Kitchawan Rd, Yorktown Heights, New York 10598, US
}
\date{\today}
\begin{abstract}
\scalebox{0.35}[0.35]{\includegraphics*[viewport=-250 20 580 770]{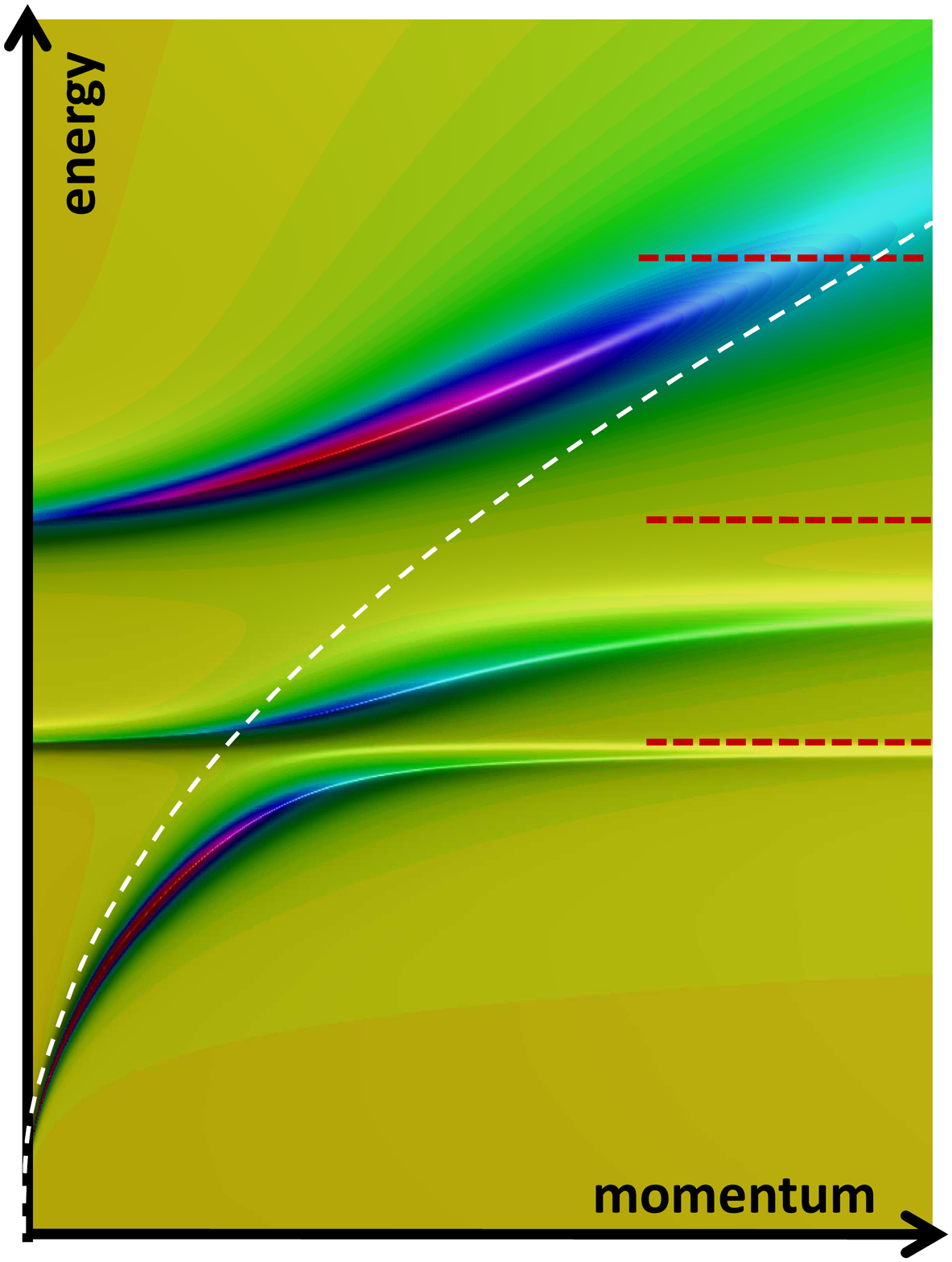}}\\
\textbf{In recent years, we have seen a rapid progress in the field of graphene plasmonics, motivated by graphene's unique electrical and optical properties, tunabilty, long-lived collective excitation and their extreme light confinement. Here, we review the basic properties of graphene plasmons; their energy dispersion, localization and propagation, plasmon-phonon hybridization, lifetimes and damping pathways. The application space of graphene plasmonics lies in the technologically significant, but relatively unexploited terahertz to mid-infrared regime. We discuss emerging and potential applications, such as modulators, notch filters, polarizers, mid-infrared photodetectors, mid-infrared vibrational spectroscopy, among many others.  \\ 
\newline}
Published in: T.Low and P.Avouris, ACS Nano, 8, p. 1086--1101 (2014)
\textbf{Glossary} Polaritons are quasiparticles resulting from the coupling of electromagnetic waves and an electric dipole-carrying excitation such as phonon and plasmon. In graphene, these excitations reside in the terahertz to mid-infrared regime. Terahertz radiations are electromagnetic waves at wavelengths from $0.1 - 1$ millimeters, while mid-infrared radiations from $3 - 8$ micrometers. The following conversion can come in handy; $1\,$eV -- $8000\,$cm$^{-1}$ -- $1.25\,$um -- $240\,$THz.
\end{abstract}
\maketitle

%\textbf{In recent years, we have seen a rapid progress in the field of graphene plasmonics, motivated by graphene's unique electrical and optical properties, tunabilty, long-lived collective excitation and their extreme light confinement. Here, we review the basic properties of graphene plasmons; their energy dispersion, localization and propagation, plasmon-phonon hybridization, lifetimes and damping pathways. The application space of graphene plasmonics lies in the technologically significant, but relatively unexploited terahertz to mid-infrared regime. We discuss emerging and potential applications, such as modulators, notch filters, polarizers, mid-infrared photodetectors, mid-infrared vibrational spectroscopy, among many others.  }
%Looking beyond the horizon, the yet unexplored general class of two-dimensional materials offers exciting opportunities in the search for better and novel plasmonic materials. We motivate this notion with graphene's immediate cousin, it's Bernal stacked bilayer, showing uniquely different plasmonic effects.

Plasmonics\cite{maier07spring} is an important subfield of photonics that deals with the excitation, manipulation, and utilization of surface plasmon polaritons (SPP, hereafter also called simply plasmons), the quantum of elementary excitation involving the collective electron oscillation.\cite{pines99} Over the last decade or so, research has focused mainly on noble metals plasmonics, with silver and gold being the predominant materials of choice. Today, metal plasmonics constitute the foundational pillars for applications as diverse as nanophotonics for integrated photonic systems,\cite{gramotnev11, novotny11, schuller10} metamaterials with unusual electromagnetic phenomena,\cite{shalaev07op,lukyanchuk10,kawata99,cai07optical} biosensing with nanostructures,\cite{kabashin09, Xu99spec} photovoltaic devices,\cite{atwater10plas} single photon transistors for quantum computing\cite{chang07, gonzalez11en} and surface-enhanced Raman spectroscopy for single molecule detection.\cite{kneipp97, nie97} These breakthroughs were realised in the visible to near-infrared frequencies.

In the ongoing search for new and better plasmonic materials,\cite{west10seach} graphene has emerged to be a very promising candidate for terahertz to mid-infrared applications,\cite{grigorenko12,koppens11nl,jablan09plas,Yan12tunable,Yan13damping,freitag13photo} the frequency range where its plasmonic resonance resides. Today, the terahertz to mid-infrared spectrum, which typically ranges from $10\,$cm$^{-1}$ to $4000\,$cm$^{-1}$, is finding a wide variety of applications\cite{tonouchi07,Ferguson02,Soref10} in information and communication, medical sciences, homeland security, military, chemical and biological sensing, spectroscopy, among many others. In this review, we describe the recent advances made in the basic understanding and engineering of plasmons in graphene metamaterials, and provide our perspective on their potential application space.
%, before extrapolating to promising future directions.   

%%%%%%%%%%%%%%%%%%%%%%%%%%%%%%%%%%%%%
\textbf{Graphene as a tunable optical material:} Accompanying the relativistic-like linear energy dispersion in graphene, with electrons traveling at a Fermi velocity only $300$ times smaller than the speed of light, are its unique electronic and optical properties.\cite{geim07, neto09rmp} Intrinsic graphene has a universal optical conductivity of $e^2/4\hbar$, where $e$ is the electronic charge and $\hbar$ is the reduced Planck constant, hence an absorption of $\alpha\pi\approx 2.3\%$ which depends only on the fine-structure constant $\alpha$.\cite{Kuzmenko08uni, mak08mea, nair08, li08dirac} Unlike metals, which have abundance of free charges, graphene is a semi-metal. Free carriers can be induced through chemical doping or electrical gating with great ease due to its two-dimensional nature. Free carriers in graphene obtained through such means can reach around $0.001-0.01$ per atom, or doping concentration of $1\times 10^{12}-1\times 10^{13}\,$cm$^{-2}$, which is significantly smaller than that of $1$ per atom for noble metals. Solid electrolyte gating can allow for higher concentration of free carriers of $0.1$ per atom, which translates to a chemical potential of $E_F\approx 1\,$eV.\cite{efetov10ph} The two-dimensionality and its semi-metallic nature of graphene therefore allows for electrical tunability not possible with conventional metals.

\begin{figure*}[htps]
\centering
\scalebox{0.55}[0.55]{\includegraphics*[viewport=0 70 800 560]{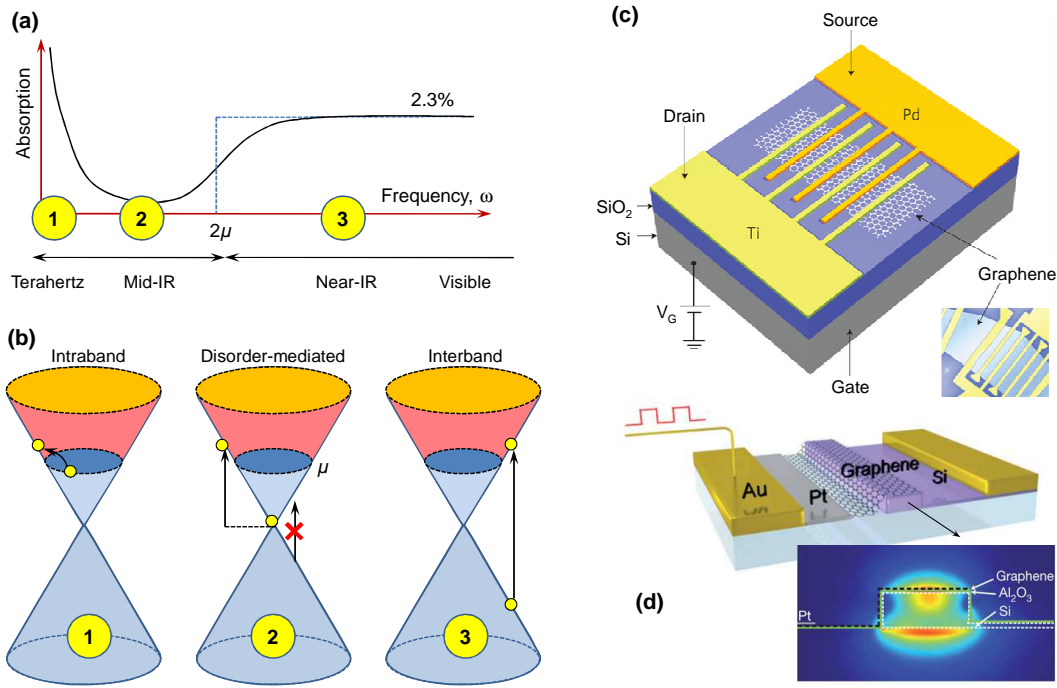}}
\caption{ \textbf{(a)} Illustration of a typical absorption spectrum of doped graphene\cite{Kuzmenko08uni, mak08mea, nair08, li08dirac}. It has characteristic features such as a Drude peak at terahertz frequencies, minimal absorption in the mid-infrared frequencies due to Pauli blocking and a transition to the universal $2.3\%$ absorption beyond the far-infrared. \textbf{(b)} Illustration of the various optical transition processes. At small $\omega$ less than the thermal energy, transitions occur via intraband processes. At finite $\omega<2\mu$, disorder plays an important role in imparting the momentum for the optical transition. A transition occurs around $\omega\approx 2\mu$, where direct interband processes lead to a universal $2.3\%$ absorption. \textbf{(c)} Schematic of a metal-graphene-metal photodetector with asymmetric metal contacts, which was operated at $10$ Gbits/s data rate with $1.55\,\mu$m light excitation as described in Ref.\,\cite{mueller10}. \textbf{(d)} Schematic of a graphene-based waveguide-integrated optical modulator reported in Ref.\,\cite{liu11mod}. Inset shows a finite element simulation of the waveguide's optical mode, designed so as to maximize the field at the interface between the waveguide and the graphene for maximal absorption efficiency. 
}
\label{fig1}
\end{figure*}

Fig.\,\ref{fig1}a-b illustrates a characteristic absorption spectrum of graphene at finite doping and the different optical transition processes involved. The spectral weight below $2E_F$ is mainly imparted to a Drude peak response at terahertz frequencies, due to intraband free carriers absorption. At near-infrared to visible frequencies, absorption is due to direct interband transitions. In doped graphene at the mid-infrared regime, Pauli-blocking occurs and the optical conductivity is minimal. The residual absorption observed in experiments at low temperature\cite{li08dirac} is generally attributed to disorder\cite{yuan11prb}. Graphene, besides being a unique material with unprecedented tunable optical properties, it is also an excellent conductor of electricity. Highest attained carrier mobility has reached $1,000,000\,$cm$^2$/Vs in suspended samples\cite{du08ballistic} and $100,000\,$cm$^2$/Vs for ultra-flat graphene on boron nitride.\cite{dean10bn,mayorov11nl} These attributes, in addition to its stability and compatibility with standard silicon processing technologies, has led to the rapid development of promising graphene-based active devices on a silicon photonic platform, such as photodetectors in optical communication data-links,\cite{xia09fast,mueller10} electro-optical modulator in silicon waveguides,\cite{liu11mod} among many others. Some of these active graphene devices are illustrated in Fig.\,\ref{fig1}c-d. In similar fashion, the utilization of graphene plasmons, a collective form of electron excitation, allows for tunable plasmonic devices.

\begin{figure}[t]
\centering
\scalebox{0.45}[0.45]{\includegraphics*[viewport=110 80 800 560]{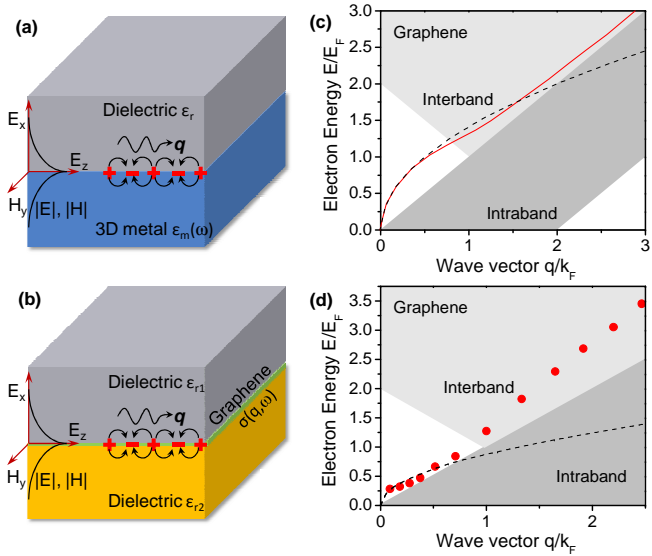}}
\caption{ \textbf{(a)} Illustration of a TM surface plasmon polariton (SPP) at a metal-dielectric interface. \textbf{(b)} Illustration of a TM plasmon mode in 2D graphene. \textbf{(c)} Plasmon dispersion in 2D graphene (solid line) calculated within the Random Phase Approximation (RPA) taken from Ref.\,\cite{hwang07dielec}. The dashed line indicates the long wavelength plasmon dispersion $\omega_{pl}\propto q$. Landau damping regions due to intraband and interband single particle electron-hole excitations are indicated. \textbf{(d)} Plasmon dispersion of 2D graphene on SiC(0001) (symbols) measured using EELS with data taken from Ref.\,\cite{liu08plas}. Dashed line is the long wavelength plasmon dispersion for reference.
}
\label{fig2}
\end{figure}

%%%%%%%%%%%%%%%%%%%%%%%%%%%%%%%%%%%%%
\textbf{Terahertz to mid-infrared plasmon-polaritons:} Conventional plasmonics emerged from the early study of SPP confined to metallic surfaces,\cite{maier05jap} where free electrons oscillate collectively in resonance with the electromagnetic field. They are confined transverse magnetic (TM) electromagnetic wave, where the electric field vector is parallel to the plasmon propagation direction i.e. its wave vector $\mathbf{q}$. This is illustrated in Fig.\,\ref{fig2}a. The dispersion relation of this SPP mode is well-known, and can be obtained by solving Maxwell's equation under appropriate boundary conditions:\cite{barnes03, maier05jap}
\begin{eqnarray}
q_{sp}=\frac{\omega}{c}\sqrt{\frac{\epsilon_r \epsilon_m(\omega)}{\epsilon_r+\epsilon_m(\omega)}}
\label{qmetal}
\end{eqnarray}
where $\epsilon_r =1$ for air. A prerequisite for its existence is that $\epsilon_r \epsilon_m(\omega)<0$, hence metals are usually used. As an example, SPP at air-silver interface has the optimal electromagnetic wave confinement in the red part of the visible spectrum\cite{jablan09plas} i.e. $\approx 18,000\,$cm$^{-1}$, with a wave-vector $q_{sp}$ an order of magnitude larger than its free space momentum, i.e. $q_0=\omega/c$. Their weaker confinement and increased losses at lower frequencies, make noble metal plasmons less appealing for applications at terahertz and mid-infrared frequencies. 

The simple dielectric-graphene-dielectric system also accommodates a TM plasmon mode, as illustrated in Fig.\,\ref{fig2}b. In the non-retarded regime where $q\gg q_0$, it has the following solution:\cite{jablan09plas, mikhailo07new, frank67pol}
\begin{eqnarray}
q\approx \frac{i2\omega \epsilon_r\epsilon_0}{\sigma(q,\omega)}
\label{qgra}
\end{eqnarray}
where $\epsilon_r$ is the average dielectric constant of its surrounding medium and $\sigma(q,\omega)$ is the non-local conductivity of graphene. In the long-wavelength (i.e. $q\ll k_F$ where $k_F$ is the Fermi wave vector) and high doping (i.e. $\hbar\omega\ll E_F$) limits, the conductivity is dominated by intraband electronic processes following the local Drude model,\cite{peres10rmp}
\begin{eqnarray}
\sigma(\omega)=\frac{ie^2|E_F|}{\pi\hbar^2(\omega+i/\tau_e)}
\label{sigmadrude}
\end{eqnarray}
Here, $\tau_e$ is the momentum relaxation time. Substituting Eq.\,\ref{sigmadrude} into Eq.\,\ref{qgra}, we obtain the plasmon dispersion relation in the long-wavelength limit,
\begin{eqnarray}
q(\omega)=\frac{2\pi\hbar^2 \epsilon_0\epsilon_r}{e^2E_F}\omega^2(1+i/\tau\omega)=\frac{\epsilon_r}{2\alpha}\frac{\omega}{\omega_F}k_0(1+i/\tau\omega)
\label{qgra2}
\end{eqnarray}
where $\alpha\equiv e^2/4\pi\hbar\epsilon_0 c  =1/137$ is the fine strucure constant and $\omega_F=E_F/\hbar$. In the mid-infrared frequency, we have $\omega/\omega_F$ of order $1$ under typical doping condition, implying a strong inherent plasmonic wave localization effect of order $1/\alpha$. Hence, the plasmon wave-vector (momentum) $q$ in graphene can be two orders larger than $q_0$, which translates to a confinement volume $10^6$ times smaller than the diffraction limit. Such extreme light confinement effect makes graphene an attractive platform for nanophotonics in these spectral ranges i.e. terahertz to mid-infrared. For clarity, the plasmon disperson $\omega_{pl}(q)$ from Eq.\,\ref{qgra2} can be obtained by setting $\tau_e=0$,
\begin{eqnarray}
\omega_{pl}=\sqrt{\frac{e^2 E_F q}{2\pi\hbar^2\epsilon_0\epsilon_r}}
\label{om_pl}
\end{eqnarray}
The $\omega_{pl}\propto\sqrt{q}$ plasmon dispersion relation is expected for all 2D systems.\cite{frank67pol} For example, using an electron concentration of $1\times 10^{13}\,$cm$^{-2}$, the plasmon frequency will reside in the mid-infrared regime of the electromagnetic spectrum. 
%In addition, with a typical mobility of $10,000\,$cm$^2$/Vs for exfoliated graphene and the momentum relaxation time related via $\tau=\mu\hbar k_F/ev_F$, we obtain a plasmon lifetime of $\approx 500\,$fs.\cite{jablan09plas}. 

%The classical electromagnetic TM mode is derived only from intraband electronic processes. However in graphene, interband transitions are also possible at frequencies $\hbar\omega>2E_F$, with a local dynamic conductivity given by\cite{falkovsky08}, 
%\begin{eqnarray}
%\nonumber
%\sigma_{inter}(\omega)=\frac{e^2}{4\hbar}\left[\frac{1}{2}+\frac{1}{\pi}\mbox{tan}^{-1}\left(\frac{\hbar\omega-2E_F}{2T}\right.\right.\mbox{...}\\
%-\left.\left.\frac{i}{2\pi}\mbox{ln}\frac{(\hbar\omega+2E_F)^2}{(\hbar\omega-2E_F)^2+(2T)^2}\right)\right]
%\label{sigmainter}
%\end{eqnarray}
%At the theshold of interband transition, $\sigma_{inter}(\omega)$ can acquire a negative imaginary contribution. As a consequence, graphene, unlike normal metals, can accomodate a new electromagnetic transverse electric (TE) mode\cite{mikhailo07new} where the electric field vector is perpendicular to the wave vector instead. This TE mode has a dispersion close to the light line, $\hbar\omega_{pl}(q)\leq cq$. Recently, a broadband graphene polarizer was implemented by coupling electromagnetic waves from an in-line optical fiber to the TE mode in graphene\cite{bao11te}. However, the SPP of TE mode is weakly localized and behaving almost like free radiation. If stronger subwavelength localization can be achieved, e.g. with nonlinear dielectric material\cite{bludovTE13}, it will be more appealing for general plasmonic applications.
%

%%%%%%%%%%%%%%%%%%%%%%%%%%%%%%%%%%%%%
\textbf{Coherent plasmons phase space: } A plasmon, the quantum of plasma oscillation, can either be self-sustaining or severely damped. For technology, we are obviously interested in the former. Landau damping is a fundamental and intrinsic damping mechanism due to the loss of collective motion to the excitation of electron-hole pairs. This occurs when the phase space of plasmon and that of intraband and interband single particle electron-hole excitations overlaps as depicted in Fig.\,\ref{fig2}c. It is also clear that the threshold for interband Landau damping can be tuned with $E_F$. 

Electron energy loss spectroscopy (EELS) is a commonly used technique for the study of electronic and plasmonic properties of materials, including graphene films. The sample is exposed to a beam of electrons with well defined kinetic energies, where the energy loss of the reflected or transmitted beam is then measured. A dominant source of energy loss is via the Coulomb interactions with the conduction electrons, including plasmons.\cite{pines52, ritchie57} This energy loss can be computed within a rigorous and microscopic theoretical framework, the Random Phase Approximation (RPA),\cite{bruus04, giuliani05, mahan00} and is defined as the imaginary part of the inverse dielectric function, i.e. $L(q,\omega)\equiv\Im[1/\epsilon_{RPA}]$. $L(q,\omega)$ is known as the electron loss function. The dielectric function, in the absence of interactions, can be computed from,
\begin{eqnarray}
\epsilon_{rpa}(q,\omega)=\epsilon_{r}-v_c \Pi_{\rho,\rho}^{0}(q,\omega)
\label{epsilon}
\end{eqnarray}
where $v_c=e^2/2q\epsilon_0$ is the 2D Coulomb interaction. $\Pi_{\rho,\rho}^{0}(q,\omega)$ is the 2D polarizability function which includes all single particle transitions between the various electronic bands. The mathematical details, including analytical form for $\Pi_{\rho,\rho}^{0}(q,\omega)$ derived at $T=0$, are described in Refs.\,\cite{hwang07dielec, wunsch06}. However, in the regime we are interested, i.e. $\omega>v_F q$ and $E_F\gg \hbar\omega$, $\Pi_{\rho,\rho}^{0}(q,\omega)$ can be approximated by,
\begin{eqnarray}
\Pi_{\rho,\rho}^{0}(q,\omega)\approx \frac{E_F q^2}{\pi\hbar^2(\omega+i/\tau_e)^2}=\frac{\epsilon_r}{v_c}\frac{\omega_{pl}^2}{\omega^2}
\label{piapprox}
\end{eqnarray}
Collective plasmonic modes can then be obtained from the zeros of the full dielectric function i.e. $\epsilon_{rpa}(q,\omega)=0$. From Eq.\,\ref{epsilon} and \ref{piapprox}, one arrived at the expected plasmon dispersion in the long-wavelength limit, $\omega=\omega_{pl}$. 

Fig.\,\ref{fig2}c shows the calculated plasmon dispersion from RPA theory using the exact form of $\Pi_{\rho,\rho}^{0}(q,\omega)$. At the long-wavelength limit, its dispersion follows the $\omega_{pl}\propto\sqrt{q}$ behavior expected for 2D plasmons.\cite{frank67pol} However, departure from this $\sqrt{q}$ behavior becomes apparent at larger $q$, where it begins to acquire a more linear dispersion. EELS experiments were performed on single and multilayer doped graphene on a SiC(0001) crystalline wafer surface and also graphene grown on Cu foil by chemical vapor deposition (CVD),\cite{liu08plas, shin11pi} where Fig.\,\ref{fig2}d shows the measured graphene plasmon dispersion. EELS is not able to resolve plasmon dispersion at small $q$. At finite $q$, we see that it disperses rather linearly instead of the $\sqrt{q}$-dependence, due to non-local effects described within the RPA. At sufficiently large plasmon momentum, i.e. $q>k_F$, the graphene plasmon enters the interband Landau damping regime, and becomes severely damped. By contrast, Landau damping for plasmons in conventional 2DEG occurs in the intraband regime instead, due to the presence of an energy gap. Note also the difference in the density dependence of $\omega_{pl}$ between graphene and 2DEG. As is apparent from Eq.\,\ref{om_pl}, $\omega_{pl}\propto n^{1/4}$ in graphene while it varies as $\omega_{pl}\propto n^{1/2}$ in conventional 2DEG. One can show this by recasting into the conventional effective mass Drude model through the simple relations, $n=k_F^2/\pi$ and $m=E_F/v_F^2$.

So far, we discussed only the coherent plasmons due to intraband processes involving the $\pi$ band of graphene. However, graphene also accommodates energetically higher plasmonic modes, such as the $\pi$ and $\pi+\sigma$ plasmons at $\approx 5\,$eV and $\approx 15\,$eV respectively\cite{gass08}. These plasmonic modes arise from the $\pi$ band and interband processes with the high energy $\sigma$ band. However, these plasmonic modes reside within the Landau-damped regions, with measured full-width at half maximum (FWHM) of the plasmon loss peak of order $1-10\,$eV.\cite{gass08, liu08plas, shin11pi} As a result, they are not of interest to the plasmonic applications we are considering and will not be discussed further in this article. 

%%%%%%%%%%%%%%%%%%%%%%%%%%%%%%%%%%%%%
\begin{figure*}[htps]
\centering
\scalebox{0.55}[0.55]{\includegraphics*[viewport=0 50 800 560]{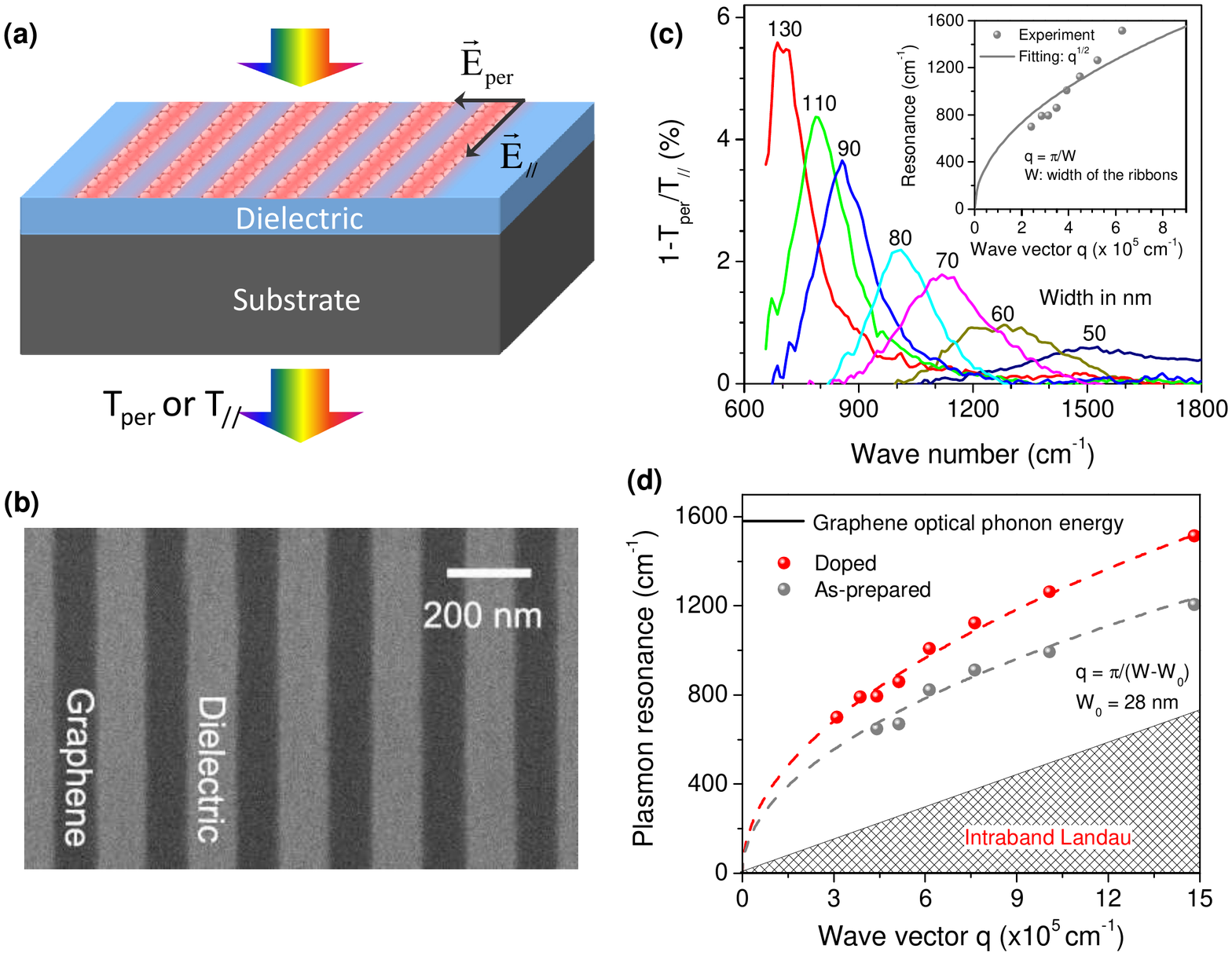}}
\caption{  \textbf{(a)} Mid-infrared transmission measurement setup for graphene nanoribbons, which also involves an infrared microscope coupled to a Fourier-transform infrared spectrometer (FTIR).  \textbf{(b)} SEM image of an array of graphene nanoribbons with width of $100\,$nm. \textbf{(c)} Extinction spectra $Z(\omega)=1-T_{per}/T_{par}$, where $T_{per}$ and $T_{par}$ are the transmission of light through the ribbon array with electric field perpendicular and parallel to the ribbons. Inset: plasmon resonance frequency as a function of wave vector $q=\pi/W$, where $W$ is the width of the nanoribbon. The grey curve is a fit according to $\omega_{pl}\propto \sqrt{q}$.  \textbf{(d)} The same plasmon resonance data (red dots) as in (c), now plotted as a function of wave-vector defined as $q=\pi/W_e$, where $W_e=W-W_0$ is an effective electrical width with $W_0=28\,$nm. Plasmon resonance data for a lower Fermi level case are also plotted (grey dots). Dashed curves are again fitting to $\omega_{pl}\propto \sqrt{q}$. Figures are taken from Ref.\,\cite{Yan13damping} with permission from Nature Publishing.
}
\label{fig3}

\centering
\scalebox{0.65}[0.65]{\includegraphics*[viewport=0 250 800 560]{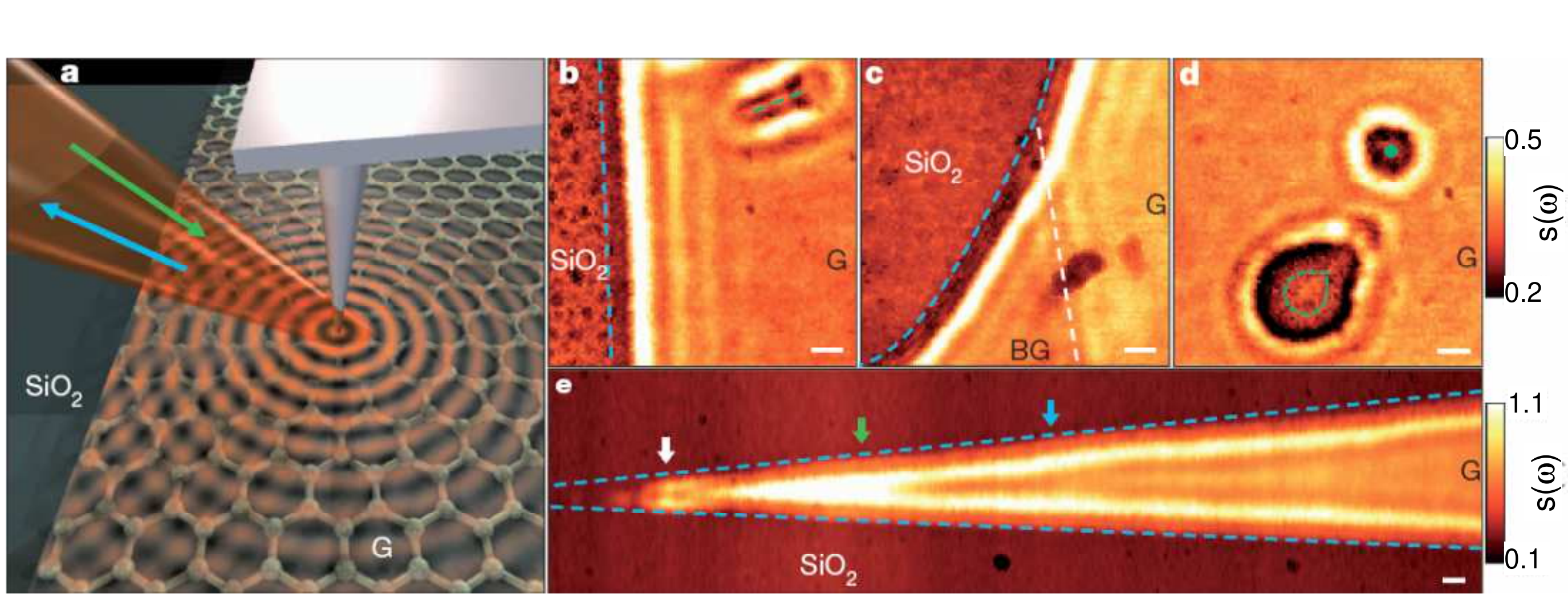}}
\caption{  \textbf{(a)} Diagram of an infrared nano-imaging experiment at the surface of graphene on SiO$_2$. Green and blue arrows display the directions of incident and back-scattered light, respectively. Concentric red circles illustrate plasmon waves launched by the illuminated tip.  \textbf{(b-e)} Images of infrared amplitude $s$ ($\omega=892\,$cm$^{-1}$) defined in the text taken at zero gate voltage. These images show a characteristic interference pattern close to graphene edges (blue dashed lines) and defects (green dashed lines and green dot), and at the boundary between single and bilayer graphene (white dashed line). Locations of boundaries and defects were determined from AFM topography taken simultaneously with the near-field data. Scale bars, $100\,$nm. All data were acquired at ambient conditions. Figures are taken from Ref.\,\cite{fei12gate} with permission from Nature Publishing.
}
\label{fig4}
\end{figure*}

\textbf{Confined plasmon-polaritons in nanostructures:} The size of a material can influence how it interacts with light especially when it is much smaller than the free space wave-length. This fact was appreciated long ago by Mie,\cite{kreibig95, mie08} when he explained the origin of the observed red color in spherical gold nanoparticles solution by simply solving Maxwell equation with the same bulk material dielectric properties, but with modified boundary conditions due to reduced size. Discontinuities in the electric permittivity can set up standing localized plasmon waves confined to the metal surfaces, when the light frequency is tuned to its plasmon resonance. In similar fashion, 2D localized plasmons can also be excited in lithographically designed thin film metal slabs, disks and their periodic arrays.\cite{berini00, braun09, dionne05, spevak09reson}

Recently, localized plasmonic modes has been experimentally observed in graphene micro- to nano-ribbons and nano-disks arrays ,\cite{ju11, Yan12tunable, Yan13damping} through prominent absorption peaks in transmission spectroscopy measurements.  In particular, polarization sensitive ribbon arrays provide a pedagogical framework for the study of basic properties of graphene plasmons,\cite{ju11, Yan13damping} as will be made apparent below. Electromagnetic simulations have shown that the plasmonic response between an individual ribbon and its arrays are indistinguishable, when the lattice constant of the array is more than twice the ribbon’s width,\cite{nikitin12sur} implying that ribbon-to-ribbon coupling can be neglected in this limit. Plasmonic resonances within an individual ribbon occur when $q\approx (2n+1)\pi/W$, where $W$ is the ribbon's width and $n=0,1,2,\ldots$. This is the condition that the plasmon half-wavelengths will be able to fit within the ribbon width. Only plasmonic modes with odd multiples of half-wavelengths\cite{mikhailov05micro} couple with light as this produces an effective charge dipole that creates the necessary restoring force for collective charge oscillations.  

Fig.\,\ref{fig3}a-b illustrates the experimental measurement scheme used to study a graphene nanoribbon array on a diamond-like carbon (DLC) substrate as described in Ref. \cite{Yan13damping}. The electromagnetic response of the array is characterized by its extinction spectrum, $Z(\omega)=1-T_{per}/T_{par}$, where $T_{per}$ and $T_{par}$ are the transmission of light through the ribbon array with electric field perpendicular and parallel to the ribbons. By excitating the localized plasmons, the extinction spectrum $Z(\omega)$ revealed prominent the resonant peaks shown in Fig.\,\ref{fig3}c at room temperature. By contrast, analogous plasmonic absorption was observed in conventional two-dimensional electron gas systems only at $4.2\,$K.\cite{allen77} The graphene plasmon dispersion $\omega_{pl}(q)$ can be mapped out by performing the measurement for ribbon arrays of different widths. Correcting for an apparent difference between the electrical and physical width of the ribbons, a good agreement between the observed localized plasmon dispersion with the classical result of Eq.\,\ref{qgra2} was found as shown in Fig.\,\ref{fig3}d. Unlike plasmonic resonances in metallic nano-particles, graphene plasmons reside in the terahertz to mid-infrared regime. In addition, the two-dimensionality and semi-metallic nature of graphene allows for electrical tunability of these plasmonic resonances.\cite{ju11,Yan13damping,chen12op, fei12gate}

%%%%%%%%%%%%%%%%%%%%%%%%%%%%%%%%%%%%%

\textbf{Seeing and launching propagating plasmons:} Previously, we discussed how far-field scattered light produced by interaction with plasmons can allow one to derive information about the plasmonic behavior, such as its dispersion. Development of scanning near-field optical microscopy (SNOM)\cite{pohl93} has provided an opportunity for direct imaging of the SPP fields in conventional metal plasmonics with nanometer range resolution, where SPP scattering, interference effects and localization have been visualized.\cite{zayats05physrep} Recently, a similar scattering-type SNOM has allowed for nano-imaging of infrared graphene plasmons\cite{chen12op, fei12gate} as illustrated in Fig.\,\ref{fig4}a. Here, the sharp tip of an atomic force microscope, with radius of curvature $a\approx 25\,$nm, is illuminated with a focused infrared beam, imparting the required momentum of order $1/a$ for the excitation of graphene plasmons. Using a free space incident light of wavelength $11.2\,\mu$m (or frequency $\omega=892\,$cm$^{-2}$), graphene plasmon in the form of a standing wave is observed. The graphene plasmon was found to have a wavelength of $200\,$nm, consistent with the estimate from Eq.\,\ref{qgra2} (here, graphene is supported on a SiO$_2$ substrate with a doping of $8\times 10^{12}\,$cm$^{-1}$), and two orders of magnitude smaller than the free space wavelength. Images of interference patterns generated by the plasmon wave close to the graphene edges, atomic defects, and at boundary of monolayer-bilayer graphene are displayed in Fig.\,\ref{fig4}b-e.

The ability for controlled launching of propagating plasmons is a key element to nanophotonics applications. Conventional plasmonics approaches include the use of a prism, either in the Otto or Kretschmann configurations,\cite{krets68, otto68} where the evanescent wave associated with total internal reflection excites propagating surface plasmons along the metal-dielectric interface. Excitation can also be performed using a diffraction grating, which shifts the wave-vector of incident electromagnetic wave to match that of the surface plasmon.\cite{maier07spring} These techniques can also be applied to graphene,\cite{bludov13} among other proposed methods such as diffractive grating etched into the underlying substrate,\cite{gao12grat,gao13grat} modulated conductivity in graphene\cite{bludov13} and via surface acoustic waves with a piezoelectric material.\cite{schiefele13} Up to date, experimental demonstration of propagating graphene plasmons via these approaches has not yet been demonstrated.

%%%%%%%%%%%%%%%%%%%%%%%%%%%%%%%%%%%%%
\begin{figure*}[htps]
\centering
\scalebox{0.65}[0.65]{\includegraphics*[viewport=0 170 800 560]{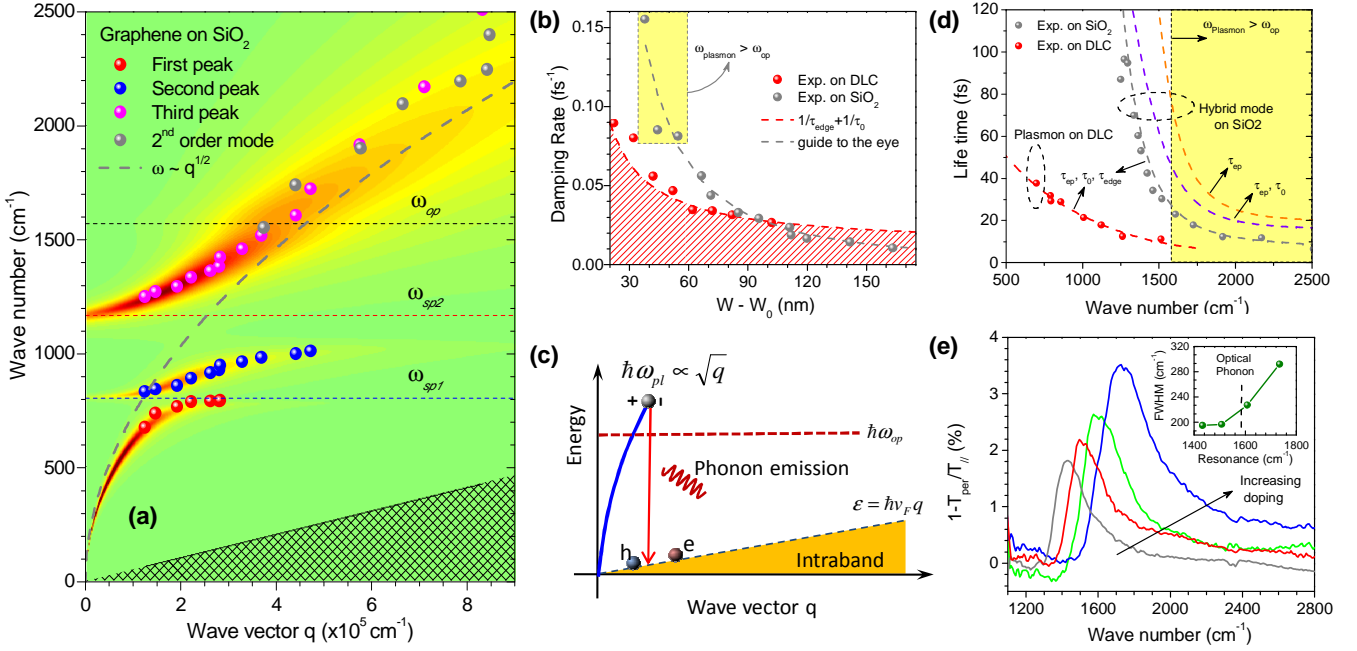}}
\caption{ \textbf{(a)} Plasmon frequency as a function of wave vector $q=\pi/W_e$ ($W_e$ is an effective electrical width) of hybrid plasmon-phonon coupled modes measured using the FTIR transmission spectroscopy technique. It is overlayed with the calculated loss function, plotted as a two-dimensional pseudo-color background. The dashed line represents the plasmon frequency without considering plasmon-phonon hybridization. Two surface polar phonons and the intrinsic optical phonon frequencies are indicated.  \textbf{(b)} Damping rates of plasmons in graphene ribbons with similar doping on DLC (red) and SiO$_2$ (grey) as a function of $W_e$. For ribbons on SiO$_2$, peak 3 in (a) is analyzed. The red dashed curve gives the calculated damping rates including edge and the DC scattering time $\tau_0$.  \textbf{(c)} Illustration of the plasmon damping process through the emission of an optical phonon, which brings it into the intraband Landau damping regime creating electron-hole pairs.  \textbf{(d)} Plasmon lifetimes of ribbons on DLC (red dots) and SiO$_2$ (grey dots) as a function of plasmon resonance frequency. Dashed curves are calculated results including various scattering processes as described in the text.  \textbf{(d)} Extinction spectra of a ribbon with width $W=100\,$nm on SiO$_2$ at 4 different doping levels. Inset shows the extracted FWHM as a function of the corresponding plasmon resonance frequency.  Figures are taken from Ref.\,\cite{Yan13damping} with permission from Nature Publishing. 
}
\label{fig5}
\end{figure*}

\textbf{Coupling plasmons with phonons:} Atomic vibrations in polar crystals can also couple resonantly with electromagnetic fields. Hybridization between plasmons and the polar optical phonon modes has long been studied in the context of doped bulk semiconductors.\cite{mooradian66} At the surfaces of these polar semiconductors, there are Fuchs-Kliewer surface optical (SO) phonons\cite{fuchs65} which have been studied extensively in conventional two-dimensional electron gas system.\cite{matz81} For isotropic polar dielectric material, the frequency of SO phonon is related to its bulk transverse and longitudinal optical (i.e. TO and LO) phonons through $\omega_{SO}=\sqrt{(\epsilon_0 +1)/(\epsilon_{\infty}+1)}\omega_{TO}$ \cite{fuchs65}, where $\epsilon_0$ ($\epsilon_{\infty}$)  is the static (high frequency) dielectric constant and $\omega_{LO}/\omega_{TO}=\sqrt{\epsilon_0/\epsilon_{\infty}}$ i.e. the Lyddane-Sachs-Teller relation. Recently, the potential benefits of coupling with these long-lived phonons for plasmonics applications have been recognized.\cite{hillenbrand02phonon, huber06, neubrech08} In addition, the surface polar phonon itself can be excited by light, forming phonon-polariton as demonstrated using infrared nanoscopy on SiC surfaces.\cite{hillenbrand02phonon}

When graphene is placed on a polar substrate, the electronic degrees of freedom in graphene, including its collective plasmon mode, can couple with these SO phonons as seen in recent experiments\cite{fei11infra, liu08plas, Yan13damping, koch10} via long-range Fr$\ddot{o}$hlich coupling, leading to a modified plasmonic dispersion.\cite{hwang10plas} The dispersion of these coupled plasmon-phonon modes were recently measured.\cite{Yan13damping} In the experiment,\cite{Yan13damping} the extinction spectra for graphene ribbon arrays on a SiO$_2$ substrate with $W$ ranging from $60- 240\,$nm were measured. There are three resonance peaks within the measured frequency range of $650-6000\,$cm$^{-1}$, due to the two relevant SO phonon modes of SiO$_2$ at $806$ and $1168\,$cm$^{-1}$. The interaction of plasmon with these phonon modes can be described within a generalized RPA theory as described below. Fig.\,\ref{fig5}a plots the measured resonances overlaid on the calculated RPA loss function with excellent agreement, where the strength of the plasmon-phonon coupling dictates the amount of anti-crossing between the plasmon and phonon modes.

Dielectric function including electron coupling with these SO phonons can be written as (see also the Supplemental Information of Ref.\,\cite{Yan13damping}):
\begin{eqnarray}
\nonumber
\epsilon_{rpa}(q,\omega)=\epsilon_{r}-v_c \Pi_{\rho,\rho}^{0}(q,\omega)-\mbox{...}\\
\sum_{SO}\frac{\epsilon_{r}\tilde{\omega}_{SO}^2}{(\omega+i/\tau_{SO})^2-\omega_{SO}^2+\tilde{\omega}_{SO}^2}
\label{epsilonsio}
\end{eqnarray}
where $\tau_{SO}$ is the lifetime of the SO phonon mode,
\begin{eqnarray}
\tilde{\omega}_{SO}&\equiv& \sqrt{\frac{4\pi}{\hbar}\omega_{SO}{\cal F}^2}\\
{\cal F}^2 &\equiv& \frac{\hbar\omega_{SO}}{2\pi}\left(\frac{1}{\epsilon_{\infty}+1}-
\frac{1}{\epsilon_{0}+1}\right)
\end{eqnarray}
${\cal F}^2$ describes the Fr$\ddot{o}$hlich coupling strength, $\epsilon_{0}$ ($\epsilon_{\infty}$) are the low (high)
frequency dielectric constant of the dielectric, which for SiO$_2$ is $3.9$ ($2.5$). The frequencies of these coupled plasmon-phonon modes can be obtained by solving for $\epsilon_{rpa}=0$ numerically. In the simple case of only one SO mode, and setting $\tau_{e}=\tau_{SO}=0$, the coupled plasmon-phonon modes reduces to a simple biquadratic equation given by,\cite{Yan13damping}
\begin{align}
\omega_{\pm}^2=\frac{\omega_{pl}^2+\omega_{SO}^2}{2}
\pm \frac{\sqrt{(\omega_{pl}^2+\omega_{SO}^2)^2-4\omega_{pl}^2(\omega_{SO}^2-\tilde{\omega}_{SO}^2)}}{2}
\end{align}
In the limit of zero coupling, i.e. $\tilde{\omega}_{SO}\rightarrow 0$, we recover the expected limits $\omega_{+}\rightarrow\omega_{SO}$ and $\omega_{-}\rightarrow\omega_{pl}$. The frequencies of these various SO phonon modes are: $\omega_{SO1}=806\,$cm$^{-1}$ and $\omega_{SO2}=1168\,$cm$^{-1}$. Typical quasiparticle lifetimes are $\tau_e=0.1\,$ps and $\tau_{SO}=1\,$ps. The coupling parameters obtained by fitting to experimental coupled plasmon-phonon resonances are ${\cal F}^2_{SO1}=0.2\,$meV and ${\cal F}^2_{SO2}=2\,$meV. Another SO phonon mode at $\omega_{SO}=460\,$cm$^{-1}$ (with  ${\cal F}^2_{SO}=0.2\,$meV\cite{low12cool}) resides outside the experimental frequency range. Hence, the choice of substrate can be used to engineer the plasmon dispersion in graphene. 

%%%%%%%%%%%%%%%%%%%%%%%%%%%%%%%%%%%%%
\textbf{Plasmon damping pathways and lifetimes: }When the plasmon enters the phase space for single particle transitions (i.e. intraband or interband), the plasmon dissociates into electron-hole pairs, losing its coherence and spectral weight. Plasmons can also decay into photons via radiative processes\cite{mikhailov96rad} which become insignificant in the electrostatic limit i.e. when the size of plasmonic structure smaller than the wavelength of light. In addition, inelastic scattering with optical phonons and other carrier scattering processes can also contribute to plasmon damping. From the measured extinction spectra described above, the plasmon damping rate $\Gamma_{pl}$ can be extracted from the resonance linewidth i.e. full-width-half-maximum (FWHM), and is related to its lifetime via $\tau_{pl}=[2\Gamma_{pl}]^{-1}$. 

Fig.\,\ref{fig5}b and d shows the measured plasmon damping rate and lifetime is a function of both the ribbon's width and the plasmon frequency for both the graphene nanoribbons array on DLC and SiO$_2$ substrate. Here, the plasmon resides outside the Landau damping regions as shown in Fig.\,\ref{fig5}a. A phenomenological model which reasonably accounts for the experimentally observed plasmon damping rate, $\Gamma_{pl}$, is given by,\cite{Yan13damping}
\begin{eqnarray}
\Gamma_{pl}(\omega,W)=\Gamma_{0}(\omega)+\frac{2v_F}{W-W_0}
\label{dampfor}
\end{eqnarray}
$\Gamma_0$ characterizes the plasmon damping processes in bulk graphene, which includes scattering via optical phonons,\cite{Yan13damping,jablan09plas} impurities,\cite{principi13dis} defects\cite{langer10} and electron-electron interactions.\cite{principi13ee} The electron momentum relaxation time, $\tau_0$, obtained from dc Hall mobility measurements or the Drude peak in ac optical conductivity provide reasonable estimates of the plasmon lifetime due to quasi-elastic processes. Electrical transport in large scale grown CVD graphene and epitaxial graphene on SiC substrate are also affected by wrinkles\cite{zhu12wrin} and atomic steps\cite{low12step} respectively, with measured electron mobility of order $\mu\approx 1,000\,$cm$^2$/Vs, which translates to $\tau_0\approx 50\,$fs. On the other hand, exfoliated graphene on boron nitride substrate can exhibits electron mobility of $\mu\approx 60,000\,$cm$^2$/Vs\cite{dean10bn}, or $\tau_0\approx 3\,$ps. 

However, highly inelastic processes due to intrinsic optical phonons, although not important at $\omega\rightarrow 0$,\cite{chen08ph} can provide significant additional energy loss channel when the plasmon energy exceeds the optical phonon emission threshold, $\omega_{pl}>\omega_{op}$. The long-wavelength optical phonons in graphene are at an energy of $\omega_{op}=0.2\,$eV, or $1580\,$cm$^{-1}$. As shown in Fig.\,\ref{fig5}d, the plasmon lifetime decays rapidly to $20\,$fs as its plasmon energy exceeds that of the optical phonon. This process is schematically illustrated in Fig.\,\ref{fig5}c, where through the emission of an optical phonon, the plasmon is brought into the intraband Landau damping regime with the creation of electron-hole pairs. Calculation of the plasmon lifetime due to this process, $\tau_{e-ph}(\omega)$, is described in detail in the Suppl. Info. of ref.\,\cite{Yan13damping} and ref.\,\cite{jablan09plas}. Evidence for the optical phonon mediated damping can be further strengthened by studying the carrier density dependence of the plasmon linewidth as shown in Fig.\,\ref{fig5}e.  As the plasmon frequency, which increases with doping, exceeds the graphene optical phonon energy, we observe a substantial increase in its resonance linewidth as shown in the inset of Fig.\,\ref{fig5}d. This observation further reinforces our picture of the plasmon decay channel via optical phonon emission.

The latter contribution in Eq.\,\ref{dampfor} is a finite size effect resulting in plasmon lifetime of $50\,$fs for $100\,$nm width nanoribbons array. Ignoring other contributions, the plasmon damping due to this finite size effect goes as $1/q$, or inversely proportional to the confinement. This general trend has been observed in experiments.\cite{tassin13sc} Analogous finite-size effects also govern plasmon damping in metallic nanoparticles.\cite{link99} In this case, the damping is proportional to $1/r$, where $r$ is the radius of the nanoparticle. There are different physical models for this behavior, many of them summarized in Ref. \cite{kreibig95}, with the dominant mechanism depending on the particular experiment. The phenomenological model, Eq.\,\ref{dampfor}, provides a satisfactory description of the experimentally observed graphene plasmon lifetime.

The role of phonons in the damping of plasmons is complex. Infrared phonons, such as the polar phonons in SiO$_2$, can couple remotely with the graphene plasmon. The resultant hybrid plasmon-phonon modes in turn inherit part of the long lifetime time of the phonon, typically of order of $1\,$ps. Analogous to a pair of classical coupled harmonic oscillators, the nature and quality of the coupled mode (that is, phonon- or plasmon-like) depends on it's resonating frequency. For example, if it is resonating at a frequency close to one of the SiO$_2$ SO phonon, it would exhibit  a narrower spectral width, inherited from the relatively long pico-second phonon lifetime. As shown in Fig.\,\ref{fig5}d, the lifetime of the hybrid plasmon-phonon mode diverges at frequency close to one of the SiO$_2$ phonon at $1168\,$cm$^{-1}$. 

%\begin{figure}[htps]
%\centering
%\scalebox{0.45}[0.45]{\includegraphics*[viewport=135 105 800 560]{fig6.pdf}}
%\caption{ \textbf{(a)} Scanning electron micrograph of a graphene disk array, with diameter $3\,\mu$m and lattice constant of $4.5\,\mu$m.  \textbf{(b)} Extinction spectrum at zero magnetic field where the solid curve is a fit to a single damped oscillator model. The deviation on the high-frequency side of the peak is due to the higher order mode. The charge distributions of the first and second order dipolar modes are illustrated in the inset. \textbf{(c)} Magnetic field dependence of the peak frequencies $\omega_{+}$ and $\omega_{-}$, their difference is also shown. Solid curves are fits based on Eq.\,\ref{gramag}. \textbf{(d)} Magnetic field dependence of the FWHM of the plasmon resonances. Figures are taken from Ref.\,\cite{yan12mag} with permission from the American Chemical Society. 
%}
%\label{fig6}
%\end{figure}

The application of magnetic field can also strongly modify the plasmon lifetime. The cyclotron spectra of graphene in a non-quantizing magnetic field, $B$, given by $\omega_c=eB/m_c$, where $m_c=E_F/v_f^2$ is the cyclotron mass, which has a finite value dependent on the carrier density.  In the absence of a magnetic field, the extinction spectra of an array of graphene microdisks measured with FTIR transmission spectroscopy technique reveals prominent resonance due to the localized 2D graphene plasmon.\cite{yan12mag} The presence of a magnetic field would exert an additional cyclotron force, splitting the resonance into  two peaks which move in opposite directions with increasing field,\cite{yan12mag} with energies given by,
\begin{eqnarray}
\omega_{\pm}=\sqrt{\omega_{pl}^2+\omega_c^2/4}\pm \omega_c/2
\label{gramag}
\end{eqnarray}
where $\omega_{pl}$ is the localized plasmon in the absence of a magnetic field. Similar behavior was also observed at natural inhomogeneities (e.g. step and wrinkles) on the graphene surface.\cite{crassee12}

The Faraday rotation spectra of these split magnetoplasmons show that each peak is excited by a different circular polarization direction. Most interestingly, the measured FWHM for $\omega_{-}$ becomes narrower while that of $\omega_+$ gets broader with increasing field.\cite{yan12mag} With increasing magnetic field, the zero field plasmon splits into a “bulk” and an “edge” mode. The latter forms skipping orbits along the edges of the disks, effectively transforming the 2D plasmon into a 1D-like plasmon. Such edge current in a broken time reversal symmetric system has the characteristic of reduced backscattering, much akin to that of quantum Hall edge current, resulting in the reduced plasmon linewidth.\cite{yan12mag} More recently, the edge plasmon dynamics in graphene disks were also measured using real time techniques.\cite{petkov13} The dramatic reduction in backscattering in the edge plasmon mode was verified and lifetimes as long as $50\,$ps (attenuation length of $70\pm30\,$mm) were observed. This is about three orders of magnitude longer than the Drude relaxation time of the sample ($0.05\,$ps for a sample mobility of $5,000\,$cm$^2$/Vs). Here, these measurements were performed under quantizing magnetic field for the $v=2$ Landau level. The above results indicate that magnetic fields provide a powerful way of tuning not only the energy, but also the lifetime of plasmons in graphene.

%%%%%%%%%%%%%%%%%%%%%%%%%%%%%%%%%%%%%

\begin{figure*}[htps]
\centering
\scalebox{0.51}[0.51]{\includegraphics*[viewport=0 100 800 560]{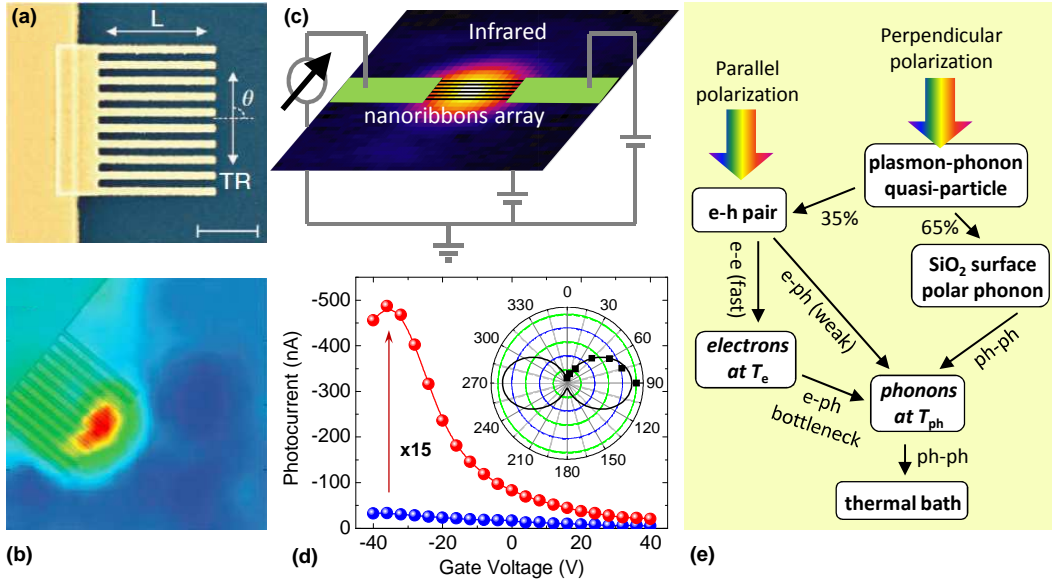}}
\caption{ \textbf{(a)} Scanning electron microscopy micrographs of the graphene devices with plasmonic nanostructures in false colors. Blue, graphene; purple, SiO$_2$ ($300\,$nm); yellow, Ti/Au electrodes. Scale bar, $20\,\mu$m. \textbf{(b)} Photovoltage maps of the device with light polarization perpendicular to the ribbons excited at $514\,$nm. \textbf{(c)} Schematic of the photoconductivity setup with a graphene nanoribbon superlattice. Infrared laser light at $10.6\,$mm is chopped at $1.1\,$KHz and the photocurrent is analysed by a lock-in amplifier referenced to the chopping frequency. \textbf{(d)} Gate-voltage-dependent photocurrent of a superlattice of $140\,$nm  graphene nanoribbon array with incident infrared light polarization perpendicular  (red spheres)  and parallel  (blue spheres) to the ribbons. Applied source-drain bias is $2\,$V. Inset shows the polarization dependence of the peak photocurrent under the same conditions. \textbf{(e)} Illustration of the mechanism of phonon and hot electron generation through decay of the hybrid plasmon-phonon quasi-particles. Perpendicular polarization excites primarily the plasmon-phonon mode, while parallel polarization excites individual electron-hole pairs. The plasmon-phonon quasi-particle decays mainly through SO phonons into other phonons, while electron-hole pairs decay primarily into hot electrons. Electrons thermalize among themselves at a temperature $T_e$ and phonons among themselves at a temperature $T_{ph}$. A bottleneck exists between electron and phonon baths, preventing full equilibration of electrons and phonons in graphene. Relevant scattering mechanisms (electron-electron, electron-phonon or phonon-phonon) are indicated. Figures are taken from Refs.\,\cite{echtermeyer11} and \cite{freitag13photo} with permission from Nature Publishing. 
}
\label{fig7}
\end{figure*}

\textbf{Enhanced optoelectronics with plasmons:} Graphene is a very unique optoelectronic material. Its large optical phonon energies ($E_{op}=200\,$meV$\,\gg kT$) and mismatch of Fermi and sound velocities ($v_F=100v_s$) imply limited scattering phase space with its intrinsic phonons, which underlies the very weak electron-phonon coupling in graphene when compared to other material systems.\cite{chen08ph,efetov10ph} Hence, upon carrier excitation with light or other means, the electronic temperature in graphene can be driven far from equilibrium providing the potential for efficient optoelectronic devices.\cite{low12cool} The optoelectronic response in graphene can be driven by bolometric,\cite{freitag12photo,yan12bi} thermoelectric,\cite{xu09ther,gabor11hot} or photovoltaic\cite{freitag12photo,mueller09con} effects, depending on the device design and operating conditions. Due to the limited absorption of 2D graphene, its response is not particularly strong, and many strategies have been proposed to overcome this limitation. For example, microcavity-induced optical confinement leads to twenty-fold enhancement in photocurrent,\cite{engel12cav,furchi12} silicon-waveguide-integrated graphene photodetectors also improves responsivities\cite{gan13chip,pospischil13} and by integrating metallic plasmonic nanostructures with graphene photodetectors as depicted in Fig.\,\ref{fig7}a-b, an order of magnitude enhancement in photo-responsivity has been achieved.\cite{echtermeyer11, liu11color} Engineering the physical dimension of metallic plasmonic nanostructures allows for spectral selective amplification of photo-response. Research efforts have focused primarily on the near-infrared to visible regions mainly for applications in optical communications.  

The terahertz and mid-infrared bands are important spectral ranges where more efficient and compact imaging and sensing devices are needed for critical security and local communications applications. Although state-of-the-art infrared detectors based on semiconductor quantum well photoconductors have response time and sensitivity approaching fundamental limits, they require cryogenic temperature operation at $4.2\,$K.\cite{rogalski10} Utilizing graphene's intrinsic plasmons would enable tunable enhanced light absorption, especially in the mid and far infrared range.\cite{li08dirac} Gate-tunable plasmonic enhanced photo-detection at room temperature in the mid-infrared regime with graphene nanoribbons array was recently demonstrated.\cite{freitag13photo} This is illustrated in Fig.\,\ref{fig7}c-d. Here, a hybrid plasmon-phonon mode is excited at $943\,$cm$^{-1}$ with a CO$_2$ laser (with perpendicular polarization), which upon decay raises both the electron and phonon temperatures ($T_e$ and $T_{ph}$). The elevated temperature then produce a change in the electrical conductivity, hence a change in the photocurrent. A schematic of the photocurrent generation mechanism is shown in Fig.\,\ref{fig7}e. Despite this being the first demonstration of optoelectronic response driven by intrinsic plasmons in graphene metamaterials, the observed photoresponse enhancement is already more than an order of magnitude under ambient conditions for a $140\,$nm nanoribbons array as shown in Fig.\,\ref{fig7}d. In addition, the narrow linewidth of $\approx\,$$100\,$meV, further allows for gate-controlled switching of the plasmonic effect.

%%%%%%%%%%%%%%%%%%%%%%%%%%%%%%%%%%%%%
\begin{figure*}[htps]
\centering
\scalebox{0.6}[0.6]{\includegraphics*[viewport=0 150 800 560]{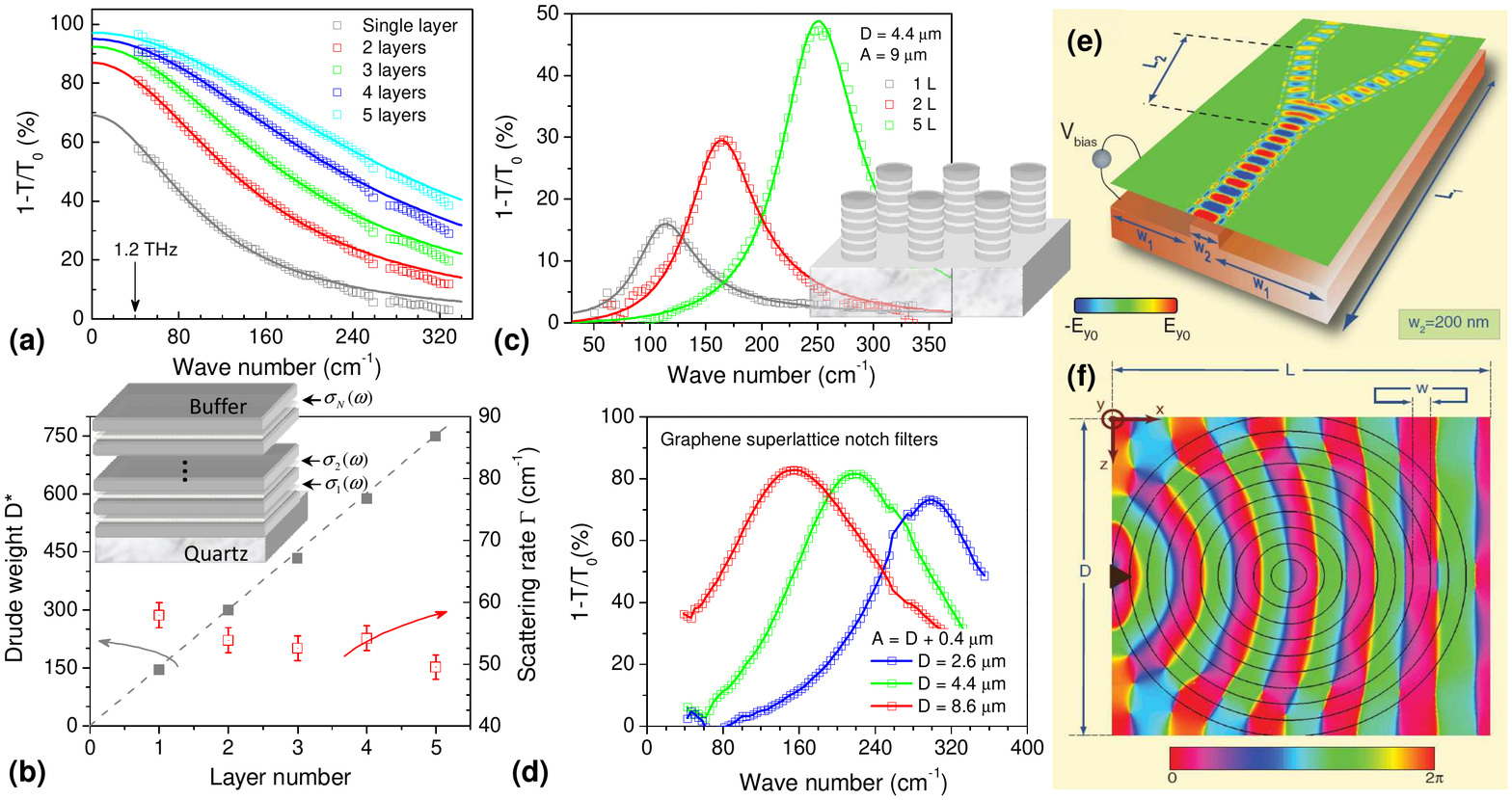}}
\caption{  \textbf{(a)} Terahertz electromagnetic wave shielding using transparent graphene/insulator stacks as illustrated in the inset of (b). Extinction spectra quantify the shielding effectiveness using stacked devices with one, two, three, four and five layers of graphene, respectively. Solid lines are fitted curves. \textbf{(b)} Fitted Drude weight and scattering width as a function of graphene layer number in the stacked devices. The Drude weight increases linearly with layer number, while the scattering width stays constant. \textbf{(c)} Plasmons in patterned graphene/insulator stacks. Extinction in transmission in stacked plasmonic devices with one, two and five graphene layers. The graphene plasmonic device is formed by patterning the stacked layers into microdisks in a triangular lattice as shown in the inset, where $d$ is the diameter of the disk and $a$ the lattice constant of the array. \textbf{(d)}  Extinction spectra of tunable terahertz filters using stacked devices with five graphene layers. The resonance frequency can be tuned by varying the diameter $d$ of the disks. In these filters, the lattice constant $a$ is always $400\,$nm larger than the disk diameter $d$. \textbf{(e)}  Guiding of plasmon in 2D graphene utilizing spatially-varying conductivity (or doping) through different capacitive coupling to the underlying back-gate. Simulation results of the electric field component $E_y$ (snap-shot in time) for an infrared-guided wave at frequency of $30\,$THz with the ribbon-like section split into two paths. \textbf{(f)} Implementation of transformation optics in 2D graphene. Here a simulation shows a 2D version of the Luneburg lens of the phase of $E_y$ at frequency of $30\,$THz. Figures (a-d) and (e-f) are taken from Refs.\,\cite{Yan12tunable} and \cite{vakil11trans} with permission from Nature Publishing and The American Association for the Advancement of Science, respectively.
}
\label{fig8}
\end{figure*}

\textbf{Active metamaterials:} Metamaterials are artificially structured materials with engineered electromagnetic properties not commonly found in nature, such as negative index-of-refraction,\cite{shalaev07op} electromagnetically-induced transparency (EIT).\cite{lukyanchuk10}, superlensing\cite{kawata99} and cloaking\cite{cai07optical} Since the electromagnetic response of metamaterials can be tailored to a particular region of the electromagnetic spectrum, especially the ``terahertz gap'' ($0.1-10\,$THz) region where most of naturally occurring materials are unresponsive, metamaterials are playing an increasingly important role in creating necessary functional devices for the rapidly developing terahertz technologies.\cite{tonouchi07} Terahertz functional devices including phase and amplitude modulators based on metallic metamaterials on Schottky diode structures\cite{chen06meta,chenmeta09} and narrow-band perfect absorbers\cite{landy08} have been demonstrated, but with limited active electrical tunability at room temperature. 

Graphene metamaterials offer a very wide design space for functional devices. Examples involve tunable graphene/insulator stacks, far-infrared notch filters with $8.2\,$dB rejection ratios and capable of shielding $97.5\%$ of the incoming terahertz radiation, while at the same time remaining relatively transparent elsewhere,\cite{Yan12tunable} as illustrated in Fig.\,\ref{fig8}a-b. Better shielding can be achieved with increasing layer number $N$, since the Drude weight scales linearly with $N$ as shown in Fig.\,\ref{fig8}b. Transparent far-infrared filters can also be implemented with similar graphene/insulator stacks in the form of micro-disk arrays as illustrated in Fig.\,\ref{fig8}c-d. Here, electromagnetic resonances can be tuned through various means. For example, the plasmon resonance increases with $1/\sqrt{d}$ where $d$ is the structural size, with the carrier density as $n^{1/4}$ and $n^{1/2}$ for single and stacked graphene layers respectively,\cite{hwang09double} with the number of stacked graphene layers as $N^{1/2}$, or via inter-disk interaction through varying the lattice constant $a$. Fig.\,\ref{fig8}c and d illustrate two particular examples of tuning, by varying $a$ and $d$, respectively. In principle, varying the doping and microstructure dimensions would enable devices with electromagnetic response ranging from terahertz to mid-infrared, operating at room temperature. Simple graphene micro- and nano-ribbons arrays have also allowed the realization of terahertz to mid-infrared linear polarizers with extinction ratio of $\approx 90\%$.\cite{ju11,Yan13damping} Integrating graphene with an array of meta-atoms and metallic wire gate electrodes, gate-controlled graphene-based metamaterials allow for modulation of both the amplitude and phase of transmitted terahertz wave by up to $47\%$ and $32.2^{o}$ respectively.\cite{leemeta12} Furthermore, the tunability of graphene provides a way to dynamically control the phase of a reflected electromagnetic field. Using an array of reflective cells, which individually produce a specific phase-shift upon reflection of the wave, would allow the construction of a desirable far-field beam, with applications as reflectarray antennas\cite{carrasco13,huang08array,hum13} e.g. for space applications.  Various theoretical proposals, yet to be experimentally demonstrated, include complete optical absorption,\cite{thon12com} graphene mantle cloak,\cite{chen11cloak} transformation optics,\cite{vakil11trans} hyperbolic metamaterials and hyperlenses based on graphene-insulator stacks.\cite{iorsh13,andrei12hyper} Fig.\,\ref{fig8}e-f shows two illustrative examples of transformation optics using graphene,\cite{vakil11trans} by designing and manipulating spatially inhomogeneous nonuniform conductivity patterns across the 2D graphene.

%%%%%%%%%%%%%%%%%%%%%%%%%%%%%%%%%%%%%

\begin{figure}[htps]
\centering
\scalebox{0.57}[0.57]{\includegraphics*[viewport=160 202 800 550]{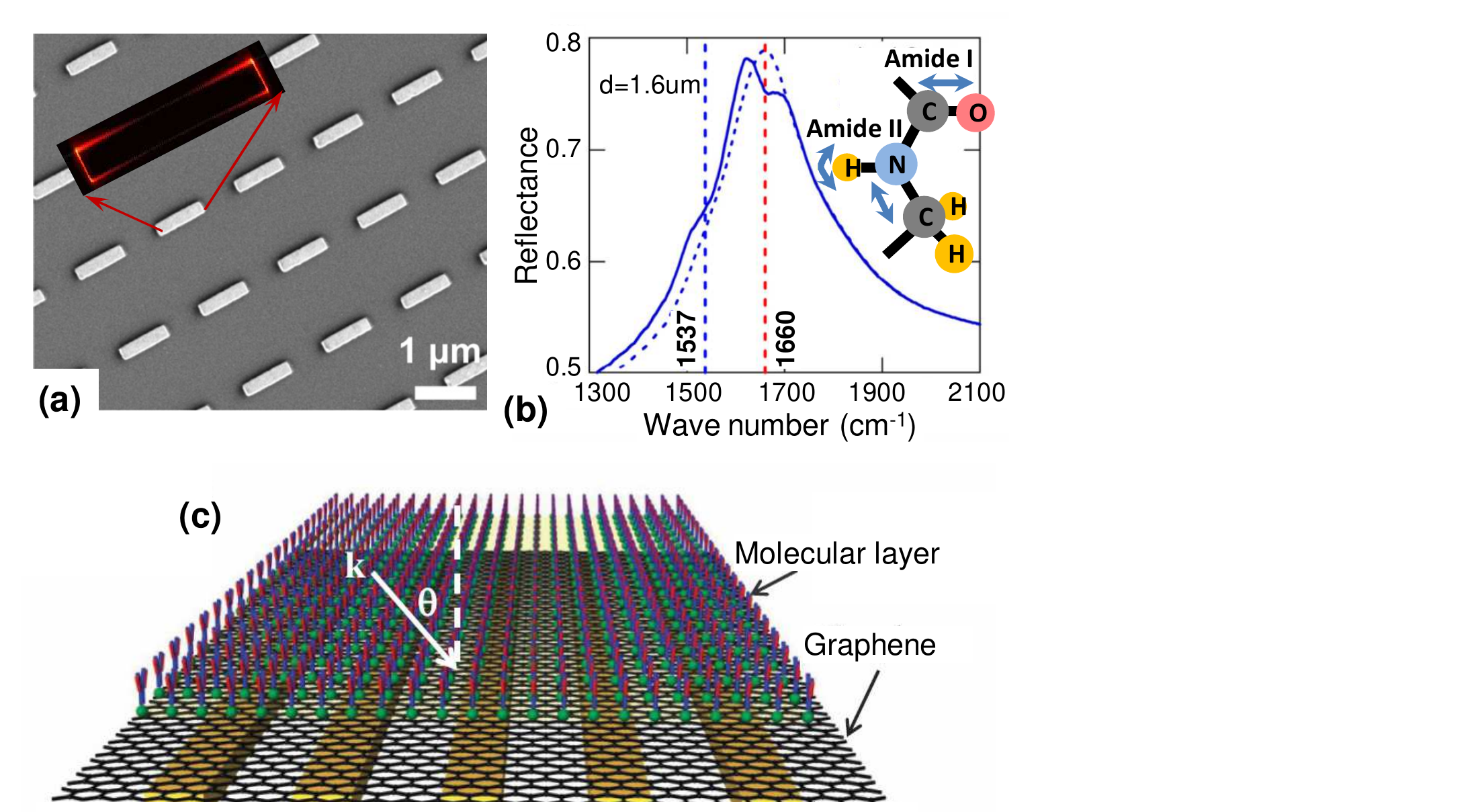}}
\caption{ \textbf{(a)} Scanning electron micrograph of a periodic array of gold nanorods designed such that their plasmonic resonance resides near the vibrational modes of a silk fibroin protein sample. Inset shows the simulated field intensity around the rod. \textbf{(b)} Reflectance spectra from a $1.6\,\mu$m periodic array before (dashed line) and after coating it with a $2\,$nm thick protein film (solid line). The amide-I and II vibrational modes of the protein backbone are illustrated in the inset. \textbf{(c)} Illustration of a graphene grating structure used for the excitation of plasmons and the detection of thin film molecular layers. Figures (a-b) and (c) are taken from Refs.\,\cite{adato09} and \cite{liu13grat} with permission from The Proceedings of the National Academy of Science and American Physical Society respectively. 
}
\label{fig9}
\end{figure}

\textbf{Mid-infrared vibrational spectroscopy:} Vibrational spectra provide fingerprints of molecular structures and as such they are used extensively in science and technology. The two techniques commonly used to obtain vibrational spectra are infrared absorption spectroscopy and Raman scattering.\cite{colthup90} Nanoscience and technology demands spectroscopy on smaller and smaller samples and even single molecules\cite{nie97} and there is increasing need for better detectivity of vibrational spectra. Plasmonic effects have served as a powerful means to enhance optical fields and are utilized in various surface-enhanced optical phenomena such as Raman scattering and second harmonic generation\cite{ponath91}. In general, the effectiveness of a plasmonic system to enhance the optical field is often expressed by its quality factor $Q$, given by $Q=-\mbox{Re}[\epsilon_m]/\mbox{Im}[\epsilon_m]$, where $\epsilon_m$ is the complex dielectric function of the metal particle. In the case of noble metal plasmonic systems, $Q$ is in the range of $10-100$, and the local optical field intensity enhancement responsible for the increase in absorption/emission rates is proportional to $Q^2$, and $Q^4$ for Raman scattering. 

In metal particles, the plasmon resonance occurs when $|\epsilon_m+\epsilon_s|$ is minimum, where $\epsilon_s$ is the dielectric function of the surrounding medium.\cite{kreibig95} For noble metals (Au and Ag), the plasmon resonances are typically in the visible part of the spectrum, while molecular vibrational mode frequencies are in the mid-infrared. These plasmonic resonances could be red-shifted up to the near-infrared by modifying the dielectric environment, e.g. by coating with a high permittivity dielectric.\cite{xu03} By switching from spherical to ellipsoidal particles, the fundamental plasmonic mode can be split into two branches, with the low-energy resonance occurring when the length of the rod is a multiple of the exciting wavelength.\cite{neubrech06rod,novotny07rod} Fig.\,\ref{fig9}a shows a periodic array of gold plasmonic nanorods designed for the detection of vibrational modes of a protein monolayer.\cite{adato09} It has a plasmonic resonance which resides close to two vibrational modes of the silk fibroin protein at $1537$ and $1660\,$cm$^{-1}$, see also inset of Fig.\,\ref{fig9}b. The measured reflectance spectra in Fig.\,\ref{fig9}b show clear signatures of these vibrational modes for just a $2\,$nm thick protein film.  In graphene nanoribbons and microribbons, localized plasmon resonances can occur in the near-infrared to far-infrared. In addition, the graphene plasmon resonance is tunable by changing the carrier density electrostatically or chemically. This invokes the interesting possibility of using patterned graphene or graphene on grating structures for mid-infrared vibrational spectroscopy of absorbed molecules. Fig.\,\ref{fig9}c illustrates a schematic of such an application.\cite{liu13grat} Below, we consider a pedagogical example which illustrates the basic principles of how plasmons in graphene can be used to enhance the infrared activity of phonons.

Consider graphene's immediate cousin: it's AB Bernal-stacked bilayer version\cite{novoselov06bi,ohta06,mcCann06}. In  monolayer graphene, the in-plane optical phonon at the $\Gamma$ point does not have any infrared activity. However, in bilayer, the $\Gamma$ phonons of the two layers can couple to form an in-phase (symmetric) and an out-of-phase (anti-symmetric) mode, where inversion symmetry is no longer invariant in the latter.\cite{Yan09ph} As a result, the anti-symmetric mode becomes infrared active. It appears as a peak in absorption absorption at $\omega$$\,\approx\,$$0.2\,$eV, with a strong dependence of the peak intensity and of its Fano-type lineshape on the applied gate voltage.\cite{tang09fano,kuzmenko09ph} In the absence of any plasmonic enhancement, the measured extinction of this infrared-active phonon mode in bilayer is about $1\%$.\cite{tang09fano,kuzmenko09ph,yan13bi} Plasmons in bilayer graphene nanostructures can enhance the phonon infrared activity, as shown in a most recent experiment,\cite{yan13bi} achieving at least a five-fold enhancement of the measured extinction.

%\begin{figure*}[htps]
%\centering
%\scalebox{0.6}[0.6]{\includegraphics*[viewport=0 327 800 560]{fig10.pdf}}
%\caption{ \textbf{(a)} 
%Real part of bulk bilayer graphene conductivity (solid line) computed at $T=300\,$K with the Fermi energy at $0.3\,$eV, constant damping of $10\,$meV and a zero band-gap. This is compared with the monolayer case. $\sigma_0$ is the universal conductivity of $e^2/4\hbar$ and $e^2/2\hbar$ for monolayer and bilayer, respectively. \textbf{(b)} RPA electron loss function $L(q,\omega)$ for bilayer graphene computed under the same conditions as in (a), assuming a background dielectric constant of $\kappa=2.5$, showing different spectra computed at different plasmon momenta $q$. Inset shows a color plot of $L(q,\omega)$ in the vicinity of the optical phonon at $0.2\,$eV. \textbf{(c)} RPA electron loss function $L(q,\omega)$ computed at a chemical potential of $\mu=0.6\,$eV with a background dielectric constant of $\kappa=1$. Green lines are the boundaries of the Landau damped regions. Here we see two plasmonic modes, a classical 2D plasmon that disperses as $\sqrt{q}$ and a high energy optical-like plasmonic mode. Figures are taken from Ref.\,\cite{low13bi} with permission from The American Physical Society. 
%}
%\label{fig10}
%\end{figure*}

The effect of electron-phonon interaction on the plasmonic response of graphene can be described within the RPA formalism, according to the charged-phonon model.\cite{cappelluti10cp,low13bi} Renormalized by many-body interactions, this `dressed' phonon gives rise to a Fano asymmetric spectral lineshape. The Fano feature develops as a result of the interference between the discrete phonon mode and the `leaky' plasmonic mode; the electronic lifetime is significantly shorter than that of the phonon, broadening the former into a quasi-continuum. At zero detuning, a very narrow transparent window emerges within the broadly opaque plasmonic absorpion.\cite{low13bi,yan13bi} This enhancement of infrared phonon spectral weight with decreasing detuning underlies the basic physics of plasmon enhanced infrared spectroscpy.\cite{aroca04} Similarly, this basic principle can also be extended to that of mid-infrared vibrational spectroscopy of absorbed molecules as illustrated in Fig.\,\ref{fig9}c. In this case, the infrared phonon (vibration) resides outside of graphene where the plasmon-phonon interaction is remote, but the physical principle is the same. Indeed, recent classical electromagnetic calculations\cite{liu13grat} of the system in Fig.\,\ref{fig9}c reveal similar type of Fano lineshape and transparency features. Other calculations suggest giant levels of light concentration leading to huge infrared-absorption enhancement in certain graphene nanostructures.\cite{Thongrattanasiri13ir}

\textbf{Concluding remarks:} In summary, there are a number of unique properties of graphene plasmonics that distinguish it from conventional metal plasmonics and many new possible areas of applications. First, the two types of plasmonics materials are fundamentally different. Graphene is the ultimate thin 2D material, covalent, flexible and inert, which can be fabricated and patterned using standard techniques such as CVD growth, lithography and etching. The energy of graphene plasmons is tunable either by electrostatic (gate) or chemical doping. Unlike noble metal plasmons whose prime application range is in the visible part of the spectrum, graphene plasmonics operate at long wavelengths: middle, far infrared and terahertz ranges. This enables infrared photonic applications (e.g. infrared photodetection, enhanced infrared absorption, optical communications) and helps to bridge the so-called THz gap. Moreover, graphene is compatible with silicon electronics and photonics and can be incorporated in these technologies. It can also be used in conjunction with conventional metamaterials and structures, such as nanoantennas, to confer to them the desired tunability, while the high sensitivity of the graphene surface plasmons to changes in the environment (changes in the dielectric function or charge transfer) suggest sensitive sensor capabilities.   

To fully exploit the possibilies offered by graphene plasmonics, a number of advances are required. First, more fundamental science work is needed to fully understand the behavior of plasmons in both monolayer and multiple layer graphenes, in particular the damping processes. The detailed behavior of these plasmons is largely determined by the quality of the graphene itself. Large scale grown high quality graphene is highly desirable in many of the applications we discussed. Advances in the high quality graphene growth and in the passivation chemistry of edges would reduce scattering and increase the Q-factors of the corresponding plasmonic devices. Development of controllable and stable chemical doping for graphene is highly desirable.  While use of localized plasmons in patterned graphene provides a very convenient way of optically accessing these excitations, edge-scattering increases their damping rates. For this reason alternative ways of exciting plasmons in extended graphene such as the use patterned substrates, e.g. gratings, should be pursued. 

At the same time, we should be reminded that graphene is only a particular example of 2D van der Waals layered materials. In the larger picture, we have other 2D materials such as boron nitride,\cite{novoselov2d05} transition metal dichalcogenides,\cite{novoselov2d05,wang12tmd,chhowalla13} metallic oxides\cite{ruben11} and many others. These materials include metals, semi-metals, semiconductors, topological insulators, common insulators, including hybrid heterostructures,\cite{novoselov2d05,wang12tmd,chhowalla13,liam13hetero,ruben11} and comes with interesting electrical and optical properties.\cite{zhang09topo,neto01,wilson75,mak12valley,zeng12valley} From this standpoint, the research on graphene plasmonics is a precursor to plasmonics with 2D materials, most of which are unexplored to-date. These diverse material properties allows for different and new application space beyond what conventional bulk metal plasmonics material can possibly offer.

%The plasmons of other 2D materials, such as metal chalcogenides, and of heterostructures are next to be explored may also offer new exciting possibilities.

%We conclude this review with a discussion of current technological challenges and trends. Large scale grown high quality graphene is highly desirable in many of the applications we discussed. Improvement in the electrical properties of these materials\cite{novoselovrm} should make possible plasmonic applications which requires smaller dissipative losses. At the same time, we should be reminded that graphene is only a particular example of 2D van der Waals layered materials. In the larger picture, we have other 2D materials such as boron nitride\cite{novoselov2d05}, transition metal dichalcogenides\cite{novoselov2d05,wang12tmd,chhowalla13}, metallic oxides\cite{ruben11} and many others. These materials include metals, semi-metals, semiconductors, topological insulators, common insulators, including hybrid heterostructures\cite{novoselov2d05,wang12tmd,chhowalla13,liam13hetero,ruben11}, and comes with interesting electrical and optical properties accompanying the new 2D materials\cite{zhang09topo,neto01,wilson75,mak12valley,zeng12valley}. From this standpoint, the research on graphene plasmonics is a precursor to plasmonics with 2D materials, most of which are unexplored to-date. These diverse material properties allows for different and new application space beyond what conventional bulk metal plasmonics material can possibly offer.

\textbf{Acknowledgement:} The authors are grateful for the fruitful collaboration and discussions with Hugen Yan, Marcus Freitag, Fengnian Xia, Francisco Guinea, Damon Farmer and Wenjuan Zhu.


\end{document}